\RequirePackage[hyphens]{url}
\RequirePackage[prologue,table]{xcolor}

\documentclass[10pt,sigconf,authorversion]{acmart}

\copyrightyear{2020}
\acmYear{2020}
\setcopyright{acmlicensed}
\acmConference[SIGMOD'20]{Proceedings of the 2020 ACM SIGMOD International Conference on Management of Data}{June 14--19, 2020}{Portland, OR, USA}
\acmBooktitle{Proceedings of the 2020 ACM SIGMOD International Conference on Management of Data (SIGMOD'20), June 14--19, 2020, Portland, OR, USA}
\acmPrice{15.00}
\acmDOI{10.1145/3318464.3386146}
\acmISBN{978-1-4503-6735-6/20/06}
\usepackage{graphicx}
\graphicspath{{./},{images/},{plots/}}

\usepackage[export]{adjustbox}
\usepackage{amsmath}
\usepackage{amssymb}
\usepackage{array}
\usepackage{booktabs} \usepackage{bm}
\usepackage{enumitem}
\usepackage{epsfig}
\usepackage{lipsum}
\usepackage{microtype}
\usepackage{pgfplots} \pgfplotsset{compat=1.14}
\usepackage{siunitx}
\usepackage{stackrel}
\usepackage{subcaption}
\usepackage{tabularx}
\usepackage{textcomp}
\usepackage{tikz}
\usepackage{todonotes}
\usepackage[normalem]{ulem}
\usepackage[table]{xcolor}
\usepackage{xspace}

\usepackage[noend,boxed]{algorithm2e}

\SetCommentSty{mycommentstyle}
\SetAlCapSkip{1ex}
\DontPrintSemicolon
\SetAlFnt{\small}

\newcolumntype{L}{>{\raggedright\arraybackslash}X}
\newcolumntype{M}[1]{>{\hsize=#1\hsize\raggedright\arraybackslash}X}\newcolumntype{C}{>{\centering\arraybackslash}X}
\newcolumntype{R}{>{\raggedleft\arraybackslash}X}
\newcolumntype{A}[1]{>{\raggedright\arraybackslash}p{#1}}
\newcolumntype{B}[1]{>{\centering\arraybackslash}p{#1}}
\newcolumntype{D}[1]{>{\raggedright\arraybackslash}m{#1}}

\newcommand{\greyrule}{\arrayrulecolor{black!30}\midrule\arrayrulecolor{black}}

\makeatletter
\def\mdseries@tt{m}
\makeatother
\usepackage{fvextra}

\usepackage[frozencache,cachedir=_minted-main]{minted}
\setminted{
  frame=single,
	linenos=false,
  escapeinside={@\{}{\}@},
	breaklines=true,
}
\setminted[json]{
	fontsize=\scriptsize,
	breakbefore={.},
	breakbeforesymbolpre={},
}
\setminted[python]{
	fontsize=\scriptsize,
	breakafter={(},
	breakaftersymbolpre={},
	breakindentnchars=4,
}

\usepackage{hyperref}
\usepackage[capitalise]{cleveref}

\newcommand{\cstyle}[1]{#1}
\newcommand{\mytt}[1]{{\small\texttt{#1}}}

\newcommand{\api}{API\xspace}
\newcommand{\autobazaar}{AutoBazaar\xspace}
\newcommand{\automl}{AutoML\xspace}
\newcommand{\btb}{\mytt{BTB}\xspace}
\newcommand{\htem}{\cstyle{pipeline hypertemplate}\xspace}  \newcommand{\hyp}{hyperparameter\xspace}
\newcommand{\Hyp}{Hyperparameter\xspace}
\newcommand{\mlb}{\mytt{MLBlocks}\xspace}
\newcommand{\mldt}{ML data type\xspace}
\newcommand{\mlp}{\mytt{MLPrimitives}\xspace}
\newcommand{\mlt}{ML task\xspace}

\newcommand{\mltt}{ML task type\xspace}

\newcommand{\Mlz}{\textit{ML Bazaar}\xspace}
\newcommand{\Orion}{Orion\xspace}
\newcommand{\piex}{\mytt{piex}\xspace}
\newcommand{\ppl}{\cstyle{pipeline}\xspace}
\newcommand{\Ppl}{\cstyle{Pipeline}\xspace}
\newcommand{\pplstep}{\cstyle{pipeline step}\xspace}
\newcommand{\Pplstep}{\cstyle{Pipeline step}\xspace}
\newcommand{\pri}{\cstyle{primitive}\xspace}
\newcommand{\Pri}{\cstyle{Primitive}\xspace}
\newcommand{\selector}{\cstyle{selector}\xspace}

\newcommand{\tem}{\cstyle{pipeline template}\xspace}

\newcommand{\tuner}{\cstyle{tuner}\xspace}

\newcommand{\son}{System 1} \newcommand{\stw}{\Mlz}
\newcommand{\sth}{System 3} \newcommand{\sfo}{System 4} \newcommand{\sfi}{System 5} \newcommand{\ssi}{System 6} \newcommand{\sse}{System 7} \newcommand{\sei}{System 8} \newcommand{\sni}{System 9} \newcommand{\ste}{System 10} 
\newcommand{\matern}{Mat\'{e}rn\xspace}

\newcommand{\ie}{i.e.\ }
\newcommand{\eg}{e.g.\ }

\newcommand{\mlpversion}{v0.2.4\xspace}
\newcommand{\nprimitivelibraries}{12\xspace}
\newcommand{\ntasktypes}{15\xspace}
\newcommand{\ndatasets}{456\xspace}
\newcommand{\npipelines}{2.5 million\xspace}  \newcommand{\nnodes}{400\xspace}

\newcommand{\timelimit}{2\xspace}

\newcommand{\pctnotsingletableclassification}{49\%\xspace}

  \newcommand{\pctimprovedbyonesd}{31.7\%\xspace}   \newcommand{\improvementinsd}{1.06\xspace}   \newcommand{\totalrfxgbpipelines}{\SI{1.86e6}\xspace}  \newcommand{\pctxgbwins}{64.9\%\xspace}
\newcommand{\ntasksxgbcomparison}{367\xspace}
\newcommand{\totalmaternsepipelines}{\SI{4.31e5}\xspace}  \newcommand{\pctsqexpwins}{60.1\%\xspace}
\newcommand{\ntasksmaterncomparison}{414\xspace}
\newcommand{\napplications}{5\xspace}

\newcommand{\mlprimitivesurl}{https://github.com/HDI-Project/MLPrimitives}
\newcommand{\mlblocksurl}{https://github.com/HDI-Project/MLBlocks}
\newcommand{\btburl}{https://github.com/HDI-Project/BTB}
\newcommand{\piexurl}{https://github.com/HDI-Project/piex}
\newcommand{\autobazaarurl}{https://github.com/HDI-Project/AutoBazaar}
\newcommand{\orionurl}{https://github.com/D3-AI/Orion}
\newcommand{\cardeaurl}{https://github.com/D3-AI/Cardea}
\newcommand{\greenguardurl}{https://github.com/D3-AI/GreenGuard}
\newcommand{\replicationurl}{https://github.com/micahjsmith/ml-bazaar-2019}
\newcommand{\mlbazaarurl}{https://mlbazaar.github.io}

\makeatletter
\if@ACM@anonymous
  \renewcommand{\mlprimitivesurl}{https://anonymous.4open.science/r/5b3f8644-d7b8-4016-a181-fee4b65c8d57/}
  \renewcommand{\mlblocksurl}{https://anonymous.4open.science/repository/c5dfbd0d-e396-411a-8f66-ac5e6d6dd577/}
  \renewcommand{\btburl}{https://anonymous.4open.science/r/e53438eb-1aed-4f9c-b5f5-ea43fa71f6af/}
  \renewcommand{\autobazaarurl}{https://anonymous.4open.science/r/99eb1c1e-580e-4556-bf86-46f943fc83d6/}
  \renewcommand{\piexurl}{https://anonymous.4open.science/r/5e0377ca-14f5-4cac-9f73-e9de80878443/}
  \renewcommand{\orionurl}{https://anonymous.4open.science/r/99eb1c1e-580e-4556-bf86-46f943fc83d6/}
  \renewcommand{\cardeaurl}{https://anonymous.4open.science/r/396ef089-0047-43c9-af92-a107c1ba180c/}
  \renewcommand{\greenguardurl}{https://anonymous.4open.science/r/be5de614-c5bc-4821-9f0c-4cf4ee38ecbc/}
  \renewcommand{\replicationurl}{https://anonymous.4open.science/r/4002e4e7-a23b-4193-b9a7-61077c6d0362/}
  \renewcommand{\mlbazaarurl}{http://mlbazaar-anonymous.s3-website.us-east-2.amazonaws.com/}
\fi
\makeatother

\DeclareMathOperator*{\argmax}{arg\,max}

\definecolor{LightCyan}{rgb}{0.88,1,1}
\definecolor{Gray}{gray}{0.8}

\setlist[itemize]{topsep=2pt,itemsep=2pt,parsep=2pt,partopsep=2pt}
\setlist[enumerate]{topsep=2pt,itemsep=2pt,parsep=2pt,partopsep=2pt}
\setlist[description]{topsep=2pt,itemsep=2pt,parsep=2pt,partopsep=2pt}

\newcommand{\mysubtitle}{Harnessing the ML Ecosystem for Effective System Development}
\newcommand{\myshortname}{The ML Bazaar}
\newcommand{\mylongname}{The Machine Learning Bazaar}

\newcommand{\mybox}[1]{\begin{tikzpicture}[font=\sffamily\scriptsize]
\node[shape=rectangle,thin,rounded corners=0.5mm,inner sep=0.3mm,text badly centered, color = black, draw=black!50!white, minimum height=1.2em, outer sep=0mm] at (0,0) {#1};
\end{tikzpicture}}

\begin{document}
\title[\myshortname: \mysubtitle]{\mylongname: \mysubtitle}

\author{Micah J. Smith}
\affiliation{\institution{MIT}}
\email{micahs@mit.edu}

\author{Carles Sala}
\affiliation{\institution{MIT}}
\email{csala@csail.mit.edu}

\author{James Max Kanter}
\affiliation{\institution{Feature Labs}}
\email{max.kanter@featurelabs.com}

\author{Kalyan Veeramachaneni}
\affiliation{\institution{MIT}}
\email{kalyanv@mit.edu}

\renewcommand{\shortauthors}{Smith et al.}

\begin{abstract}

As machine learning is applied more widely, data scientists often struggle to find or create end-to-end machine learning systems for specific tasks. The proliferation of libraries and frameworks and the complexity of the tasks have led to the emergence of ``pipeline jungles'' --- brittle, ad hoc ML systems. To address these problems, we introduce the \textit{Machine Learning Bazaar}, a new framework for developing machine learning and automated machine learning software systems. First, we introduce ML primitives, a unified \api and specification for data processing and ML components from different software libraries. Next, we compose primitives into usable ML pipelines, abstracting away glue code, data flow, and data storage. We further pair these pipelines with a hierarchy of \automl strategies --- Bayesian optimization and bandit learning. We use these components to create a general-purpose, multi-task, end-to-end \automl system that provides solutions to a variety of data modalities (image, text, graph, tabular, relational, etc.) and problem types (classification, regression, anomaly detection, graph matching, etc.). We demonstrate 5 real-world use cases and 2 case studies of our approach. Finally, we present an evaluation suite of \ndatasets real-world ML tasks and describe the characteristics of \npipelines \ppl{s} searched over this task suite.

\end{abstract}



\maketitle

\section{Introduction}
\label{sec:intro}

Once limited to conventional commercial applications, machine learning (ML) is now being widely applied in physical and social sciences, in policy and government, and in a variety of industries. This diversification has led to difficulties in actually creating and deploying real-world systems, as key functionality becomes fragmented across ML-specific or domain-specific software libraries created by independent communities. In addition, the process of building problem-specific end-to-end systems continues to be marked by ML and data management challenges, such as formulating achievable learning problems \cite{Kanter2016label}, managing and cleaning data and metadata \cite{miao2017modelhub,VanderWeide2017,bhardwaj15datahub}, scaling tuning procedures \cite{falkner2018bohb,li2018massively}, and deploying models and serving predictions \cite{Baylor2017TFXAT,crankshaw2015missing}. In practice, engineers and data scientists often spend significant effort developing ad hoc programs for new problems: writing ``glue code'' to connect components from different software libraries, processing different forms of raw input, and interfacing with external systems. These steps are tedious and error-prone and lead to the emergence of brittle ``pipeline jungles'' \cite{sculley2015hidden}.

These points raise the question, \textit{``How can we make building ML systems easier in practical settings?''} A new approach is needed to designing and developing software systems that solve specific ML tasks. Such an approach should address a wide variety of input data modalities, such as images, text, audio, signals, tables, graphs, and learning problem types, such as regression, classification, clustering, anomaly detection, community detection, graph matching; it should cover the intermediate stages involved, such as data preprocessing, munging, featurization, modeling, and evaluation; and it should enable \automl functionality to fine-tune solutions, such as \hyp tuning and algorithm selection. Moreover, it should offer coherent \api{s}, fast iteration on ideas, and easy integration of new ML innovations. In sum, this ambitious goal would allow almost all end-to-end learning problems to be solved or built using a single framework.

\begin{figure*}[!ht]
    \centering
    \includegraphics[width=1.0\linewidth]{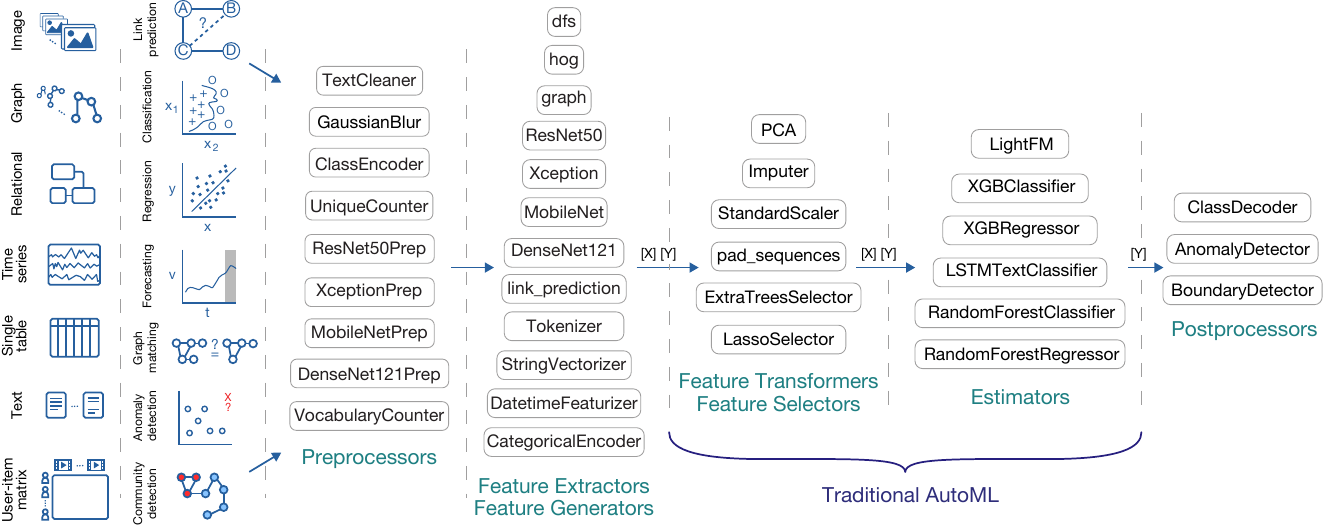}
    \caption{Various \mltt{s} that can be solved in \Mlz using composition of ML \pri{s} (abbreviated here from fully-qualified names). Primitives are categorized into preprocessors, feature processors, estimators, and postprocessors and are drawn from many different ML libraries, such as scikit-learn, Keras, OpenCV, and NetworkX, as well as custom implementations. Many additional \pri{s} and \ppl{s} are available in our curated catalog.}
    \label{fig:full_pipeline}
\end{figure*}

To address these challenges, we present the \textit{Machine Learning Bazaar},\footnote{Just as one open-source community was described as ``a great babbling bazaar of different agendas and approaches'' \cite{cathedralbazaar}, our framework is characterized by the availability of many compatible alternatives, a wide variety of libraries and custom solutions, a space for new contributions, and more.} a framework for designing and developing ML and \automl systems. We organize the ML ecosystem into composable software components, ranging from basic building blocks like individual classifiers to full \automl systems. With our design, a user specifies a task, provides a raw dataset, and either composes an end-to-end \ppl out of pre-existing, annotated, ML \pri{s} or requests a curated \ppl for their task (\Cref{fig:full_pipeline}). The resulting \ppl{s} can be easily evaluated and deployed across a variety of software and hardware settings and tuned using a hierarchy of \automl approaches. Using our own framework, we have created an \automl system which we have entered in participation in DARPA's Data-Driven Discovery of Models (D3M) program; ours is the first end-to-end, modular, publicly released system designed to meet the program's goal.

\begin{figure*}[t]
  \begin{subfigure}[b]{0.42\textwidth}
    \centering
\begin{minted}{python}
primitives = [
 'mlp.custom.ts_preprocessing.time_segments_average',
 'sklearn.impute.SimpleImputer',
 'sklearn.preprocessing.MinMaxScaler',
 'mlp.custom.ts_preprocessing.rolling_window_sequences',
 'keras.Sequential.LSTMTimeSeriesRegressor',
 'mlp.custom.ts_anomalies.regression_errors',
 'mlp.custom.ts_anomalies.find_anomalies',
]
\end{minted}
    \vspace{5px}
\begin{minted}{python}
options = {
 'init_params': {
  'mlp.custom.ts_preprocessing.time_segments_average#1':
   {'time_column': 'timestamp', 'interval': 21600},
  'sklearn.preprocessing.MinMaxScaler#1':
   {'feature_range': [-1, 1]},
  'mlp.custom.ts_preprocessing.rolling_window_sequences#1':
   {'target_column': 0, 'window_size': 250},
  'keras.Sequential.LSTMTimeSeriesRegressor':
   {'epochs': 35},
 },
 'input_names': {
  'mlp.custom.ts_anomalies.find_anomalies#1':
   {'index': 'target_index'},
 },
 'output_names': {
  'keras.Sequential.LSTMTimeSeriesRegressor#1':
   {'y': 'y_hat'},
 },
}
\end{minted}
    \caption{Python representation.}
    \label{lis:orion-pipeline}
  \end{subfigure}  \begin{subfigure}[b]{0.23\textwidth}
    \centering
    \includegraphics[clip=true,trim={78 0 10 72},width=0.85\linewidth]{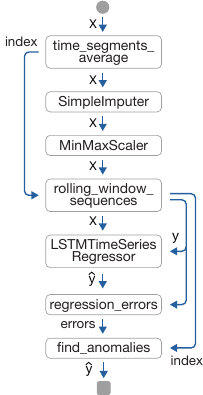}
    \vspace{40px}
    \caption{Graph representation.}
    \label{fig:orion-graph}
  \end{subfigure}  \begin{subfigure}[b]{0.32\textwidth}
    \centering
\begin{minted}{python}
from mlblocks import MLPipeline
from orion.data import load_signal
from orion.pipelines.lstm_dt import primitives, options

train = load_signal('S-1-train')
test = load_signal('S-1-test')

ppl = MLPipeline(primitives, **options)
ppl.fit(train)
anomalies = ppl.predict(test)
\end{minted}
    \vspace{80px}
    \caption{Usage with Python SDK.}
    \label{lis:orion-usage}
  \end{subfigure}
  \caption{Representation and usage of the \Orion \ppl for anomaly detection using the \Mlz framework. ML system developers or researchers describe the \ppl in a short Python snippet by a sequence of \pri{s} annotated from several libraries (and optional additional parameters). Our framework compiles this into a graph representation (\Cref{sec:mlbazaar:building-ml-pipelines:pipeline-description-interface}) by consulting meta-information associated with the underlying \pri{s} (\Cref{sec:mlbazaar:mlprimitives:annotations}). Developers can then use our Python SDK to train the \ppl on ``normal'' signals and then identify anomalies in test signals. The \mytt{MLPipeline} provides a familiar interface but enables more general data engineering and ML processing. It also can expose the entire underlying hyperparameter configuration space for tuning by our \automl libraries or others (\Cref{sec:automl}).}
  \label{fig:orion}
\end{figure*}

To preview the potential of development using our framework, we highlight the \Orion project within MIT for ML-based anomaly detection in satellite telemetry (\Cref{fig:orion}), as one of several successful real-world applications that uses \Mlz for effective ML system development (\Cref{sec:applications:use-cases}). The \Orion \ppl processes a telemetry signal using several custom preprocessors, an LSTM predictor, and a dynamic thresholding postprocessor to identify anomalies. The entire \ppl can be represented in a short Python snippet, custom processing steps are easily implemented as modular components, two external libraries are integrated without glue code, and the \ppl can be tuned using composable \automl functionality.

Our contributions in this paper include:

\begin{itemize}[leftmargin=0pt,itemindent=0pt]
\renewcommand{\labelitemi}{}

    \item \textbf{A composable framework for representing and developing ML and \automl systems}: Our framework enables users to specify a pipeline for any ML task, ranging from image classification to graph matching, through a unified API (\Cref{sec:mlbazaar,sec:automl}).

    \item \textbf{The first general-purpose automated machine learning system}: Our system, \autobazaar, is, to the best of our knowledge, the first open-source, publicly-available, system with the ability to reliably compose end-to-end, automatically-tuned, solutions for \ntasktypes data modalities and problem types (\Cref{sec:automl:building-an-automl-system}).

    \item \textbf{Successful applications}: We describe \napplications successful application of our framework on real-world problems (\Cref{sec:applications}).

    \item \textbf{A comprehensive evaluation}: We evaluated our \automl system against a suite of \ndatasets ML tasks/datasets covering \ntasktypes \mltt{s}, analyzing \npipelines scored pipelines (\Cref{sec:evaluation}).

    \item \textbf{Open-source libraries:} Components of our framework have been released as the open-source libraries \mlp, \\
    \mlb, \btb, \piex, and \autobazaar.

\end{itemize}

\section{The Machine Learning Bazaar}
\label{sec:mlbazaar}

The \Mlz is a composable framework for developing ML and \automl systems based on a hierarchical organization of and unified API for the ecosystem of ML software and algorithms. One can use curated or custom software components for every aspect of the practical ML process, from featurizers for relational datasets to signal processing transformers to neural networks to pre-trained embeddings. From these {\it\pri{s}}, data scientists can easily and efficiently construct ML solutions for a variety of \mltt{s}, and ultimately, automate much of the work of tuning these models.

\subsection{ML Primitives}
\label{sec:mlbazaar:ml-primitives}

A {\it\pri} is a reusable, self-contained, software component for ML paired with the structured \textit{annotation} of its metadata. It has a well-defined \mytt{fit}/\mytt{produce} interface wherein it receives input data in one of several formats or types, performs computations, and returns the data in another format or type. With this categorization and abstraction, widely varying functionality required to construct ML pipelines can be collected in a single location. \Pri{s} can be re-used in chained computations while minimizing glue code written by callers. An example primitive annotation is shown in \Cref{lis:primitive-example}.

Primitives encapsulate different types of functionality. Many have a learning component, such as a random forest classifier. Many \pri{s}, categorized as transformers, may have no learning component and only have a \mytt{produce} method, but are very important nonetheless. For example, the Hilbert and Hadamard transforms from signal processing would be important \pri{s} to include when building an ML system to solve a task in Internet-of-Things.

Some primitives do not change the values in the data, but simply prepare or reshape the data. These \textit{glue primitives} are intended to reduce glue code required to connect primitives into a full system. An example of this type of primitive is \mytt{pandas.DataFrame.unstack}.

\subsubsection{Annotations}
\label{sec:mlbazaar:mlprimitives:annotations}

Each \pri is annotated with machine-readable metadata that enables its usage and automatic integration within an execution engine. Annotations allow us to unify a variety of primitives from disparate libraries, reduce the need for glue code, and provide information about the tunable hyperparameters. This full annotation\footnote{The primitive annotation specification is described and documented in full in the associated \mlp library.} is provided in a single JSON file and has three major sections:

\begin{listing}[ht]
\begin{minted}{json}
{
  "name": "cv2.GaussianBlur",
  "contributors": [
    "Carles Sala <csala@csail.mit.edu>"
  ],
  "description": "Blur an image using a Gaussian filter.",
  "classifiers":
    {"type": "preprocessor", "subtype": "transformer"},
  "modalities": ["image"],
  "primitive": "mlprimitives.adapters.cv2.GaussianBlur",
  "produce": {
    "args": [{"name": "X", "type": "ndarray"}],
    "output": [{"name": "X", "type": "ndarray"}]
  },
  "hyperparameters": {
    "fixed": {
      "ksize_width": {"type": "int", "default": 3},
      "ksize_height": {"type": "int", "default": 3},
      "sigma_x": {"float", "default": 0},
      "sigma_y": {"float", "default": 0}
    },
    "tunable": {}
  },
  "...": "..."
}
\end{minted}
\caption{JSON representation of the \texttt{GaussianBlur} transformer \pri using \mlp. (Some fields are abbreviated or elided.) This \pri does not annotate any \textit{tunable \hyp{s}} but such a section marks hyperparameter types, defaults, and feasible values.}
\label{lis:primitive-example}
\end{listing}

\begin{itemize}

  \item \textit{Meta-information.} This section has the name of the primitive, the fully-qualified name of the underlying implementation as a Python object, and other detailed metadata, such as the author, description, documentation URL, categorization, and the data modalities it is most used for. This information enables searching and indexing primitives.

  \item \textit{Information required for execution.} This section has the names of the methods pertaining to \mytt{fit}/\mytt{produce} in the underlying implementation as well as the data types of the \pri{'s} inputs and outputs. When applicable, for each \pri, we annotate the \textit{\mldt{s}} of declared inputs and outputs, i.e., recurring objects in ML that have a well-defined semantic meaning, such as a feature matrix $X$, a target vector $y$, or a space of class labels \mytt{classes}. We provide a mapping between \mldt{s} and synonyms used by specific libraries as necessary. This logical structure will help dramatically decrease the amount of glue code developers must write (\Cref{sec:mlbazaar:building-ml-pipelines:steps-and-pipelines}).

  \item \textit{Information about \hyp{s}.} The third section details all the \hyp{s} of the \pri~--- their names, descriptions, data types, ranges, and whether they are \textit{fixed} or \textit{tunable}. It also captures any conditional dependencies between the \hyp{s}.

\end{itemize}

We have developed the open-source \mlp\footnote{\url{\mlprimitivesurl}} library which contains a number of primitives adapted from different libraries (\Cref{tab:mlp-catalog}). For libraries that already provide a \mytt{fit}/\mytt{produce} interface or similar (\eg scikit-learn), a \pri developer has to write the JSON specification and point to the underlying estimator class.

To support integration of \pri{s} from libraries that need significant adaptation to the \mytt{fit}/\mytt{produce} interface, \\ \mlp also provides a powerful set of adapter modules that assist in wrapping common patterns. These adapter modules then allow us to integrate many functionalities as \pri{s} from the library without having to write a separate object for each --- thus requiring us to write only an annotation file for each primitive. Keras is an example of such a library.

\begin{table}[th]
  \centering
  \begin{tabular}{llll}
    \toprule
    Source                & Count & Source         & Count \\
    \cmidrule(lr){1-2} \cmidrule(lr){3-4}
    scikit-learn          & 39    & XGBoost        & 2 \\
    MLPrimitives (custom) & 27    & LightFM        & 1 \\
    Keras                 & 25    & OpenCV         & 1 \\
    pandas                & 16    & python-louvain & 1 \\
    Featuretools          & 4     & scikit-image   & 1 \\
    NumPy                 & 3     & statsmodels    & 1 \\
    NetworkX              & 2     &                & \\
    \bottomrule
  \end{tabular}
  \caption{\Pri{s} in the curated catalog of \mlp, by library source. Catalogs maintained by individual projects may contain more \pri{s}.}
  \label{tab:mlp-catalog}
\end{table}

For developers, domain experts, and researchers, \mlp enables easy \textit{contribution} of new \pri{s} in several ways by providing \pri templates, example annotations, and detailed tutorials and documentation. We also provide procedures to validate proposed \pri{s} against the formal specification and a unit test suite. Finally, contributors can also write custom primitives.

Currently, \mlp maintains a curated \textit{catalog} of high-quality, useful \pri{s} from \nprimitivelibraries libraries,\footnote{As of \mlp \mlpversion.} as well as custom \pri{s} that we have created (\Cref{tab:mlp-catalog}).
Each \pri is identified by a fully-qualified name to differentiate \pri{s} across catalogs. The JSON annotations can then be mined for additional insights.

\subsubsection{Designing for contributions}
\label{sec:mlbazaar:mlprimitives:designing-for-contributions}

We considered multiple alternatives to the \pri{s} API, such as representing all of them as Python data structures or classes, regardless of their type (\ie transformers or estimators). One disadvantage of these alternatives is that it makes it more difficult for domain experts to contribute \pri{s}. We have found that domain experts, such as engineers and scientists in the satellite industry, prefer writing functions rather than other constructs such as classes, and many domain-specific processing methods are simply transformers without a learning component.

\subsubsection{Lightweight integration}
\label{sec:mlbazaar:mlprimitives:lightweight-integration}

Another option we considered was to enforce that every \pri --- whether brought over from a library with a compatible \api or otherwise --- be integrated via a Python class with wrapper methods. We opted against this approach as it led to excessive wrapper code and created redundancy, which made it more difficult to write \pri{s}. Instead, for libraries that are compatible, our design requires that we only create the annotation file.

\subsubsection{Language independence}
\label{sec:mlbazaar:mlprimitives:language-independence}

In this work, we focus on the wealth of ML functionality that exists in the Python ecosystem. Through \Mlz{'s} careful design, we could also support other common languages in data science like R, MATLAB, and Julia and enable multi-language \ppl{s}. Starting from our JSON \pri annotation format, a multi-language \ppl execution backend would be built that uses language-specific kernels or containers and relies on an interoperable data format such as Apache Arrow. A language-independent format like JSON provides several additional advantages. It is both machine- and human- readable and writeable. It is also a natural format for storage and querying in NoSQL document stores, allowing developers to easily query a knowledge base of primitives for the subset appropriate for a specific \mltt, for example.

\subsection{Building ML pipelines}
\label{sec:mlbazaar:building-ml-pipelines}

To solve practical learning problems, we must be able to instantiate and compose \pri{s} into usable programs. These programs must be easy to specify with a natural interface, such that developers can easily compose \pri{s} without sacrificing flexibility. We aim to support both end-users trying to build an ML solution for their specific problem who may not be savvy about software engineering, as well as system developers wrapping individual ML solutions in \automl components. In addition, we provide an abstracted execution layer, such that learning, data flow, data storage, and deployment are handled automatically by various configurable and pluggable backends. As one realization of these ideas, we have implemented \mlb,\footnote{\url{\mlblocksurl}} a library for composing, training, and deploying end-to-end ML \ppl{s}.

\begin{listing}[H]
\begin{minted}{python}
from mlblocks import MLPipeline
from mlblocks.datasets import load_umls
from mlblocks.discovery import load_pipeline

dataset = load_umls()
X_train, X_test, y_train, y_test = dataset.get_splits(1)
graph = dataset.graph
node_columns = ['source', 'target']

ppl = MLPipeline(load_pipeline('graph.link_prediction.nx.xgb'))
ppl.fit(X_train, y_train, graph=graph, node_columns=node_columns)
y_pred = ppl.predict(X_test, graph=graph, node_columns=node_columns)
\end{minted}
\caption{Usage of the \mlb library for a graph link prediction task. Curated \ppl{s} in the \mlp library can be easily loaded. \Ppl{s} provide a familiar API but enable more general data engineering and ML.}
\label{lis:ml-pipeline-api-3}
\end{listing}

\subsubsection{Steps and Pipelines}
\label{sec:mlbazaar:building-ml-pipelines:steps-and-pipelines}

We introduce {\it ML \ppl{s}}, which collect multiple \pri{s} into a single computational graph. Each \pri in the graph is instantiated in a {\it\pplstep}, which loads and interprets the underlying \pri and provides a common interface to run a step in a larger program.

We define a {\it\ppl} as a directed acyclic multigraph \\
${L = \langle V, E, \lambda \rangle}$, where $V$ is a collection of \pplstep{s}, $E$ are the directed edges between steps representing data flow, and $\lambda$ is a joint \hyp vector for the underlying \pri{s}. A valid \ppl\ --- and its generalizations (\Cref{sec:automl:templates-and-hypertemplates}) --- must also satisfy acceptability constraints that require the inputs to each step to be satisfied by the outputs of another step connected by a directed edge.

The term ``\ppl'' is used in the literature to refer to a ML-specific sequence of operations, and sometimes abused (as we do here) to refer to a more general computational graph or analysis. In our conception, we bring foundational data processing operations of raw inputs into this scope, like featurization of graphs, multi-table relational data, time series, text, and images, as well as simple data transforms, like encoding integer or string targets. This gives our \ppl{s} a greatly expanded role, providing solutions to any \mltt and spanning the entire ML process beginning with the raw dataset.

\subsubsection{\Ppl description interface}
\label{sec:mlbazaar:building-ml-pipelines:pipeline-description-interface}

\begin{figure}[th]
  \begin{subfigure}[b]{0.4\linewidth}
\begin{minted}{python}
primitives = [
  'UniqueCounter',
  'TextCleaner',
  'VocabularyCounter',
  'Tokenizer',
  'SequencePadder',
  'LSTMTextClassifier',
]
\end{minted}
    \vspace{15px}
    \caption{Python representation.}
    \label{lis:text-classifier-pipeline}
  \end{subfigure}  \begin{subfigure}[b]{0.55\linewidth}
    \centering
    \includegraphics[width=0.9\linewidth]{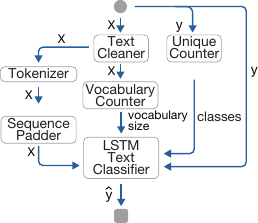}
    \caption{Graph representation.}
    \label{fig:text-classifier-graph}
  \end{subfigure}
  \caption{Recovery of ML computational graph from \ppl description for a text classification \ppl. The \mldt{s} that enable extraction of the graph, and stand for data flow, are labeled along edges.}
  \label{fig:text-classifier}
\end{figure}

Large graph-structured workloads can be difficult to specify for end-users due to the complexity of the data structure and such workloads are an active area of research in data management. In \Mlz, we consider three aspects of \ppl representation: ease of composition, readability, and computational issues. First, we prioritize easily composing complex ML \ppl{s} by providing a \textit{\ppl description interface} (PDI) in which developers specify only the topological ordering of all \pplstep{s} in the \ppl without requiring any explicit dependency declarations. These steps can be passed to our libraries as Python data structures or loaded from JSON files. Full training-time (\mytt{fit}) and inference-time (\mytt{produce}) computational graphs can then be recovered (\Cref{alg:pdi-to-graph}). This is made possible by the meta-information provided in the \pri annotations, in particular, the \mldt{s} of the \pri inputs and outputs. We leverage the observation that steps that modify the same \mldt can be grouped into the same subpath. In cases where this information does not uniquely identify a graph, the user can additionally provide an \textit{input-output map} which serves to explicitly add edges to the graph, as well as other parameters to customize the \ppl.

Though it may be more difficult to read and understand these \ppl{s} from the PDI alone as the edges are not shown nor labeled, it is easy to accompany them with the recovered graph representation (\Cref{fig:orion,fig:text-classifier}).

\begin{algorithm}[!htb]
  \KwIn{pipeline description $S = (v_1, \dots, v_n)$, source node $v_0$, sink node $v_{n+1}$}
  \KwOut{directed acyclic multigraph $\langle V, E \rangle$}
  \BlankLine
  \Begin{
      $S \leftarrow v_0 \cup S \cup v_{n+1}$ \;
      $V \leftarrow \varnothing$, $E \leftarrow \varnothing$ \;
      $U \leftarrow \varnothing$ \tcp*{unsatisfied inputs}
      \While{$S \ne \varnothing$}{
          $v \leftarrow \text{popright}(S)$ \tcp*{last \pplstep remaining}
          $M \leftarrow \text{popmatches}(U, \text{outputs}(v))$ \;
          \uIf{$M \ne \varnothing$}{
              $V \leftarrow V \cup \{v\}$ \;
              \For{$(v', \sigma) \in M$}{
                  $E \leftarrow E \cup \{(v, v', \sigma)\}$
              }
              \For(\tcp*[f]{unsatisfied inputs of $v$}){$\sigma \in  \text{\normalfont{inputs}}(v)$}{
              $U \leftarrow U \cup \{(v, \sigma)\}$ \;
              }
          }
          \Else(\tcp*[f]{isolated node}){
              \Return{\tt{INVALID}}
          }
      }
      \If(\tcp*[f]{unsatisfied inputs remain}){$U \ne \varnothing$}{
          \Return{\tt{INVALID}}
      }
      \Return{$\langle V, E \rangle$}
  }
  \caption{Pipeline-Graph Recovery. \Pplstep{s} are added to the graph in reverse order and edges are iteratively added when the step under consideration produces an output that is required by an existing step. Exactly one graph is recovered if a valid graph exists. In cases where multiple graphs have the same topological ordering, the user can additionally provide an \textit{input-output map} (which modifies the result of inputs($v$)/outputs($v$) above) to explicitly add edges and thereby select from among several possible graphs.}
  \label{alg:pdi-to-graph}
\end{algorithm}

The resulting graphs describe abstract computational workloads, but we must be able to actually execute them for purposes of learning and inference. From the recovered graphs, we could re-purpose many existing data engineering systems as backends for scheduling and executing the workloads \cite{dask,spark,palkar2018evaluating}. In our \mlb execution engine, a collection of objects and a metadata tracker in a key-value store are iteratively transformed through sequential processing of \pplstep{s}. The \Orion \ppl would be executed using \mlb as shown in \Cref{lis:orion-usage}.

\subsection{Discussion}
\label{sec:mlbazaar:discussion}

\paragraph{Why not scikit-learn?}

Several alternatives exist to our new ML \ppl abstraction (\Cref{sec:mlbazaar:building-ml-pipelines}), such as scikit-learn's \mytt{Pipeline} \cite{sklearn_api}. Ultimately, while our \ppl is inspired by these alternatives, it aims to provide more general data engineering and ML functionality. While the scikit-learn pipeline sequentially applies a list of transformers to $X$ and $y$ only before outputting a prediction, our \ppl supports general computational graphs, accepts multiple data modalities as input simultaneously, produces multiple outputs, manages evolving metadata, and can use software from outside the scikit-learn ecosystem/design paradigm. For example, we can use our \ppl to construct entity sets \cite{Kanter2015deep} from multi-table relational data for input to other \pplstep{s}. We can also support \ppl{s} in an unsupervised learning paradigm, such as in \Orion, where we create the target $y$ ``on-the-fly'' (\Cref{fig:text-classifier}).

\paragraph{Where'd the glue go?}

To connect learning components from different libraries with incompatible \api{s}, data scientists end up writing ``glue code''. Typically, this glue code is written within \ppl bodies. In \Mlz, we mitigate the need for this glue by pushing the need of \api adaptation down to the level of \pri annotations, which are written once and reside in central locations, amortizing the adaptation cost. Moreover, the need for glue code arises in creating intermediate outputs and shaping of the data. We created a number of primitives that support these common programming patterns and miscellaneous needs in development of a ML pipeline. These are, for example, data reshaping primitives like \mytt{pandas.DataFrame.unstack},  data preparation primitives like \mytt{pad\_sequences} required for Keras-based LSTMs, and utilities like \mytt{UniqueCounter} that count the number of unique classes.

\paragraph{Interactive development}

Interactivity is an important aspect of data science development for beginners and experts alike, as they build understanding of the data and iterate on different modeling ideas. In \Mlz, the level of interactivity possible depends on the specific runtime library. For example, our \mlb library supports interactive development in a shell or notebook environment by allowing the inspection of intermediate \ppl outputs and by allowing \ppl{s} to be iteratively expanded starting from a loaded \ppl description. Alternatively, ML \pri{s} could be used as a backend \ppl representation for software that provides more advanced interactivity such as drag-and-drop. For interfaces that require low latency \ppl scoring to provide user feedback such as \cite{crotty2015vizdom}, \Mlz{'s} performance depends mainly on the underlying \pri implementations (\Cref{sec:evaluation}).

\paragraph{Supporting new task types}

While \Mlz handles \ntasktypes \mltt{s} (\Cref{tab:task-type-count-template}), there are many more task types for which we do not currently provide \ppl{s} in our default catalog (\Cref{sec:evaluation:expressiveness}). To extend our approach to support new task types, it is generally sufficient to write several new \pri annotations for pre-processing input and post-processing output --- no changes are needed to the core \Mlz software libraries such as \mlp and \mlb. For example, for the anomaly detection task type from the \Orion project, several new simple \pri{s} were implemented: \mytt{rolling\_window\_sequences}, \mytt{regression\_errors}, and \mytt{find\_anomalies}. Indeed, support for a certain task type is predicated on the availability of a \ppl for that task type rather than any characteristics of our software libraries.

\paragraph{\Pri versioning}

The default catalog of \pri{s} from the \mlp library is versioned together, and library conflicts are resolved manually by maintainers through carefully specifying minimum and maximum dependencies. This strategy ensures that the default catalog can always be used, even if there are incompatible updates to the underlying libraries.
Automated tools can be integrated to aid both end-users and maintainers in understanding potential conflicts and safely bumping library-wide versions.

\section{\automl System Design and Architecture}
\label{sec:automl}

From the components of the \Mlz, data scientists can easily and effectively build ML pipelines with fixed \hyp{s} for their specific problems. To improve the performance of these solutions, we introduce the more general {\it \tem{s}} and {\it \htem{s}} and then present the design and implementation of \automl \pri{s} which facilitate \hyp tuning and model selection, either using our own library for Bayesian optimization or external \automl libraries. Finally, we describe \autobazaar, one specific \automl system we have built on top of these components.

\subsection{Pipeline templates and hypertemplates}
\label{sec:automl:templates-and-hypertemplates}

Frequently, \ppl{s} require \hyp{s} to be specified at several places. Unless these values are fixed at annotation-time, \hyp{s} must be exposed in a machine-friendly interface. This motivates \tem{s} and \htem{s}, which generalize \ppl{s} by allowing a hierarchical tunable \hyp configuration space and provide first-class tuning support.

We define a {\it\tem} as a directed acyclic multigraph ${T = \langle V, E, \Lambda \rangle}$, where $\Lambda$ is the joint \hyp configuration space for the underlying \pri{s}. By providing values $\lambda \in \Lambda$ for the unset \hyp{s} of a \tem, a specific \ppl is created.

In some cases, certain values of \hyp{s} can affect the domains of other \hyp{s}. For example, the type of kernel for a support vector machine results in different kernel \hyp{s}, and preprocessors used to adjust for class imbalance can affect the training procedure of a downstream classifier. We call these \textit{conditional \hyp{s}}, and accommodate them with \htem{s}.

We define a {\it\htem} as a directed acyclic multigraph ${H = \langle V, E, \bigcup_j \Lambda_j \rangle}$, where $V$ is a collection of \pplstep{s}, $E$ are directed edges between steps, and $\Lambda_j$ is the \hyp configuration space for \tem $T_j$. A number of \tem{s} can be derived from one \htem by fixing the conditional \hyp{s}.

\subsection{\automl Primitives}
\label{sec:automl:automl-primitives}

Just as \pri{s} units components of ML computation, \automl \pri{s} represent components of an \automl system. We separate \automl \pri{s} into {\it\tuner{s}} and {\it\selector{s}}. In our extensible \automl library for developing \automl systems, \btb,\footnote{\url{\btburl}} we provide various instances of these \automl \pri{s}.

\subsubsection{Tuners}
\label{sec:automl:automl-primitives:tuners}

Given a \tem, an \automl system must find a specific \ppl with fully-specified \hyp values to maximize some utility. Given \tem $T$ and a function $f$ that assigns a performance score to \ppl $L_\lambda$ with \hyp{s} $\lambda \in \Lambda$, the tuning problem is defined as $\lambda^* = \argmax_{\lambda \in \Lambda} f(L_{\lambda})$. We introduce {\it\tuner}{s}, \automl \pri{s} which provide a \mytt{record}/\mytt{propose} interface in which evaluation results are recorded to the \tuner by the user or by an \automl controller and new \hyp{s} are proposed in return.

\Hyp tuning is widely studied and its effective use is instrumental to maximizing the performance of ML systems \cite{bergstra2011algorithms,bergstra2012random,feurer2015efficient,snoek2012practical}. One widely used approach to hyperparameter tuning is Bayesian optimization (BO), a black-box optimization technique in which expensive evaluations of $f$ are kept to a minimum by forming and updating a meta-model for $f$. At each iteration, the next \hyp configuration to try is chosen according to an acquisition function. We structure these meta-models and acquisition functions as separate BO-specific \automl \pri{s} that can be combined together to form a \tuner. Researchers have argued for different formulations of meta-models and acquisition functions \cite{oh2018bock,wang2017new,snoek2012practical}.
In our \btb library for \automl, we implement the \mytt{GP-EI} \tuner, which uses a Gaussian Process meta-model \pri and an Expected Improvement (EI) acquisition function \pri, among several other tuners. Many other tuning paradigms exist, such as those based on evolutionary strategies \cite{loshchilov2016cma,Olson2016EvaluationOA}, adaptive execution \cite{jamieson2016nonstochastic,li2017hyperband}, meta-learning \cite{gomes2012combining}, or reinforcement learning \cite{drori2018alphad3m}. Though we have not provided implementations of these in \btb, one could do so using our common API.

\subsubsection{Selectors}
\label{sec:automl:automl-primitives:selectors}

For many \mltt{s}, there may be multiple \tem{s} or \htem{s} available, each with their own tunable \hyp{s}. The aim is to balance the exploration-exploitation tradeoff while selecting promising \tem{s} to tune. For a set of \tem{s} $\mathcal{T}$, we define the selection problem as $T^{*} = \argmax_{T \in \mathcal{T}} \, \max_{\lambda_T \in \Lambda_T} f(L_{\lambda_T})$. We introduce {\it\selector}{s}, \automl \pri{s} which provide a \mytt{compute\_rewards}/\mytt{select} \api.

Algorithm selection is often treated as a multi-armed bandit problem where the score returned from a selected template can be assumed to come from an unknown underlying probability distribution. In \btb, we implement the \mytt{UCB1} \selector, which uses the upper confidence bound method \cite{auer2002finite}, among several other \selector{s}. Users or \automl controllers can use \selector{s} and \tuner{s} together to perform joint algorithm selection and hyperparameter tuning.

\subsection{Building an \automl system}
\label{sec:automl:building-an-automl-system}

Using the ML Bazaar framework, we have built \autobazaar,\footnote{\url{\autobazaarurl}} an open-source, end-to-end, general-purpose, multi-task, automated machine learning system. It consists of several components: an \automl controller; a \ppl execution engine; data stores for metadata and \ppl evaluation results; loaders and configuration for \mlt{s}, \pri{s}, etc.; a Python language client; and a command-line interface. \autobazaar is an open-source variant of the \automl system we have developed for the DARPA D3M program.

We focus here on the core \ppl search and evaluation algorithms (\Cref{alg:autobazaar-search}). The input to the search is a computational budget and an \mlt, which consists of the raw data and task and dataset metadata --- dataset resources, problem type, dataset partition specifications, and an evaluation procedure for scoring. Based on these inputs, \autobazaar searches through its catalog of \pri{s} and \tem{s} for the most suitable \ppl that it can build.
First, the controller loads the train and test dataset partitions, $\mathcal{D}^{(train)}$ and $\mathcal{D}^{(test)}$, following the metadata specifications. Next, it loads from its default catalog and the user's custom catalog a collection of candidate \tem{s} suitable for the \mltt. Using the \btb library, it initializes a \mytt{UCB1} \selector and a collection of \mytt{GP-EI} \tuner{s} for joint algorithm selection and hyperparameter tuning. The search process begins and continues for as long as the computation budget has not been exhausted. In each iteration, the \selector is queried to select a template, the corresponding \tuner is queried to propose a \hyp configuration, a pipeline is generated and scored using cross validation over $\mathcal{D}^{(train)}$, and the score is reported back to the \selector and \tuner. The best overall \ppl found during the search, $L^*$, is re-fit on $\mathcal{D}^{(train)}$ and scored over $\mathcal{D}^{(test)}$. Its specification is returned to the user alongside the score obtained, $s^*$.

\begin{algorithm}[!htb]
\KwIn{task $t = (M, f, \mathcal{D}^{(train)}, \mathcal{D}^{(test)})$, budget $B$}
\KwOut{best \ppl $L^*$, best score $s^*$}
\BlankLine
\Begin{
      $\mathcal{T} \leftarrow \text{load\_available\_templates}(M)$ \;
  A $\leftarrow$ init\_automl($\mathcal{T}$) \tcp*{bookkeeping}
  \BlankLine
  $s^* \leftarrow +\infty, L^* \leftarrow \varnothing$ \;
  \While{$B > 0$}{
    $T \leftarrow \text{select}(A)$ \tcp*{uses \texttt{selector.select}}
    $\lambda \leftarrow \text{propose}(A, T)$ \tcp*{uses $T$'s \texttt{tuner.propose}}
    $L \leftarrow (T, \lambda)$ \;
    $s \leftarrow \text{cross\_validate\_score}(f, L, \mathcal{D}^{(train)})$ \;
    record(A, $L, s$) \tcp*{update selector and tuners}
    \If{$s < s^*$}{
      $s^* \leftarrow s$, $L^* \leftarrow L$ \;
    }
    decrease($B$) \;
  }
  \BlankLine
  $s^* \leftarrow \text{fit\_and\_score}(f, L^*, \mathcal{D}^{(train)}, \mathcal{D}^{(test)})$ \;
  \Return{$L^*, s^*$}
}
\caption{Search and evaluation of \ppl{s} in \autobazaar. Detailed task metadata $M$ is used by the system to load relevant \tem{s} and scorer function $f$ is used to score \ppl{s}.}
\label{alg:autobazaar-search}
\end{algorithm}

\section{Applications}
\label{sec:applications}

In this paper, we claim that \Mlz makes it easier to develop ML systems. We provide evidence for this claim in this section by describing \napplications real-world use cases in which \Mlz is currently used to create both ML and \automl systems. Through these industrial applications we examine the following questions: Does \Mlz support the needs of ML system developers? If not, how easy was it to extend?

\subsection{Use cases}
\label{sec:applications:use-cases}

\paragraph{Anomaly detection for satellite telemetry}
\label{sec:applications:use-cases:orion}

\Mlz is used by a communications satellite operator which provides video and data connectivity globally. This company wanted to monitor more than 10,000 telemetry signals from their satellites and identify anomalies, which might indicate a looming failure severely affecting the satellite's coverage. This time series/anomaly detection task was not initially supported by any of the \ppl{s} in our curated catalog. Our collaborators were able to easily implement a recently developed end-to-end anomaly detection method \cite{hundman2018detecting} using pre-existing transformation \pri{s} in \Mlz and by adding several new primitives: a \pri for the specific LSTM architecture used in the paper and new time series anomaly detection postprocessing \pri{s}, which take as input a time series and time series forecast, and produce as output a list of anomalies, identified by intervals $\{[t_i, t_{i+1}]\}$. This design enabled rapid experimentation through substituting different time series forecasting \pri{s} and comparing the results. In current work, they apply ML \ppl{s} to 82 publicly available satellite telemetry signals from NASA and evaluate the anomaly detections against 105 known anomalies. The work has been released as the open-source \Orion project\footnote{\url{\orionurl}} and is currently under active development.

\paragraph{Predicting clinical outcomes from electronic health records}
\label{sec:applications:use-cases:cardea}

Cardea is an open-source, automated framework for predictive modeling in health care on electronic health records following the FHIR schema. Its developers formulated a number of prediction problems including predicting length of hospital stay, missed appointments, and hospital readmission. All tasks in Cardea are multitable regression or classification. From \Mlz, Cardea uses the \mytt{featuretools.dfs} \pri to automatically engineer features for this highly-relational data and multiple other \pri{s} for classification and regression. The framework also presents examples on a publicly available patient no-show prediction problem. The framework has been released as an open-source project.\footnote{\url{\cardeaurl}}

\paragraph{Failure prediction in wind turbines}
\label{sec:applications:use-cases:greenguard}

\Mlz is also used by a multinational energy utility to predict critical failures and stoppages in their wind turbines. Most prediction problems here pertain to the time series classification \mltt. \Mlz has several time series classification \ppl{s} available in its catalog and they enable usage of time series from 140 turbines to develop multiple \ppl{s}, tune them, and produce prediction results. Multiple outcomes are predicted, ranging from stoppage and pitch failure to less common issues, such as gearbox failure. This library is released as the open-source GreenGuard project.\footnote{\url{\greenguardurl}}

\paragraph{Leaks and crack detection in water distribution systems}
\label{sec:applications:use-cases:water}

A global water technology provider uses \Mlz for a variety of ML needs, ranging from image classification for detecting leaks from images, to crack detection from time series data, to demand forecasting using water meter data. \Mlz provides a unified framework for these disparate needs. The team also builds custom \pri{s} internally and uses them directly with the \mlb backend.

\subsection{DARPA D3M program}
\label{sec:applications:d3m}

DARPA's Data-Driven Discovery of Models (D3M) program, of which we are participants, aims to spur development of automated systems for model discovery for use by non-experts. Among other goals, participants aim to design and implement \automl systems that can produce solutions to arbitrary ML tasks without any human involvement. We used \Mlz to create an \automl system to be evaluated against other teams from US academic institutions. Participants include ourselves (MIT), CMU, UC Berkeley, Brown, Stanford, TAMU, and others. Our system relies on \automl \pri{s} (\Cref{sec:automl}) and other features of our framework, but does \textit{not} use our \pri and \ppl implementations (neither \mlp nor \mlb).

We present results comparing our system against other teams in the program. DARPA organizes an evaluation every 6 months (Winter and Summer). During evaluation, \automl systems submitted by participants are run by DARPA on 95 tasks spanning several task types for three hours per task. At the end of the run, the best \ppl identified by the \automl system is evaluated on held-out test data. Results are also compared against two independently-developed expert baselines (MIT Lincoln Laboratory and Exline).

Results from one such evaluation from Spring 2018 were presented by \cite{shang2019democratizing}. We make comparisons from the Summer 2019 evaluation, the results of which were released in August 2019 --- the most recent evaluation as of this writing. \Cref{tab:darpa-summer} compares our \automl system against 9 other teams. Given the same tasks and same machine learning primitives, this comparison highlights the efficacy of the \automl primitives (\btb) in \Mlz only --- it does not provide any evaluation of our other libraries. In its implementation, our system uses a \texttt{GP-MAX} tuner and a \texttt{UCB1} selector. Across all metrics, our system places 2nd out of the 10 teams.

\begin{table}
  \footnotesize
  \begin{tabular}{lcccc}
    \toprule
    System & Top pipeline & Beats Expert 1 & Beats Expert 2 & Rank\tabularnewline
    \cmidrule(lr){1-1}\cmidrule(lr){2-4}\cmidrule(lr){5-5}
    \son & 29 & 57 & 31 & 1\tabularnewline
    \rowcolor{LightCyan} \stw & 18 & 56 & 28 & 2\tabularnewline
    \sth & 15 & 47 & 22 & 3\tabularnewline
    \sfo & 14 & 46 & 21 & 4\tabularnewline
    \sfi & 10 & 42 & 14 & 5\tabularnewline
    \rowcolor{Gray} \ssi & 8 & 43 & 15 & 6\tabularnewline
    \rowcolor{Gray} \sse & 8 & 33 & 12 & 7\tabularnewline
    \sei & 6 & 24 & 11 & 8\tabularnewline
    \sni & 4 & 25 & 13 & 9\tabularnewline
    \ste & 2 & 27 & 12 & 10\tabularnewline
    \bottomrule
  \end{tabular}
  \caption{Results from the DARPA D3M Summer 2019 evaluation (most recent conducted). Entries represent the number of ML tasks. ``Top pipeline'' is the number of tasks for which a system created a winning pipeline. ``Beats Expert 1'' and ``Beats Expert 2'' are the number of tasks for which a system beat the two expert team baselines. We highlight Systems 6 and 7 as they belong to the same teams as \cite{shang2019democratizing} and \cite{drori2018alphad3m}, respectively. (We are unable to comment on other systems as they have not yet provided public reports.) Rank is given based on number of top pipeline lines produced. The top 4 teams are consistent in their ranking even if a different column is chosen.}
  \label{tab:darpa-summer}
\end{table}

\subsection{Discussion}
\label{sec:applications:discussion}

Through these applications using the components of the \Mlz, several advantages surfaced.

\subsubsection{Composability}

One important aspect of \Mlz is that it does not restrict the user to use a single monolithic system, rather users can pick and choose parts of the framework they want to use. For example, \Orion uses only \mlp/\mlb, Cardea uses \mlp but integrates the hyperopt library for hyperparameter tuning, our D3M \automl system submission mainly uses \automl \pri{s} and \btb, and \autobazaar uses every component.

\subsubsection{Focus on infrastructure}

The ease of developing ML systems for the task at hand freed up time for teams to think through and design a comprehensive ML infrastructure. In the case of \Orion and GreenGuard, this led to the development of a database that catalogues the metadata from every ML experiment run using \Mlz. This had several positive effects: it allowed for easy sharing between team members, and it allowed the company to transfer the knowledge of what worked from one system to another system. For example, the satellite company plans to use the pipelines that worked on a previous generation of the satellites on the newer ones from the beginning. With multiple entities finding uses for such a database, creation of such infrastructure could be templatized.

\subsubsection{Multiple use cases}

Our framework allowed the water technology company to solve many different \mltt{s} using the same framework and API.

\subsubsection{Fast experimentation}

Once a baseline pipeline has been designed to solve a problem, we notice that users can quickly shift focus to developing and improving primitives that are responsible for learning.

\subsubsection{Production ready}

A fitted pipeline maintains all the learned parameters as well as all the data manipulations. A user is able to serialize the pipeline and load it into production. This reduces the development-to-production lifecycle.

\section{Experimental evaluation}
\label{sec:evaluation}

In this section, we experimentally evaluate \Mlz along several dimension. We also leverage our evaluation results to perform several case studies in which we show how a general-purpose evaluation setting can be used to assess the value of specific ML and \automl \pri{s}.

\subsection{ML task suite}
\label{sec:evaluation:ml-task-suite}

The \Mlz Task Suite is a comprehensive corpus of tasks and datasets to be used for evaluation, experimentation, and diagnostics. It consists of \ndatasets \mlt{s} spanning \ntasktypes task types. Tasks, which encompass raw datasets and annotated task descriptions, are assembled from a variety of sources, including MIT Lincoln Laboratory, Kaggle, OpenML, Quandl, and Crowdflower. We created train/test splits and organized the folder structure. Other than this, we do not do any preprocessing (sampling, outlier detection, imputation, featurization, scaling, encoding, etc.), presenting data in its raw form as it would be ingested by end-to-end \ppl{s}. Our publicly-available task suite can be browsed online\footnote{\url{\mlbazaarurl}} or through \piex,\footnote{\url{\piexurl}} our library for exploration and meta-analysis of \mlt{s} and \ppl{s}.
The covered task types are shown in \Cref{tab:task-type-count-template} and a summary of the tasks is shown in \Cref{tab:ml-task-suite-summary}.

\begin{table}
  \footnotesize
  \centering
  \begin{tabular}{lrrrrr}
  \toprule
  {} &  min &   p25 &   p50 &    p75 &       max \\
  \cmidrule(lr){2-6}
  Number of examples & 7 & 202 & 599 & 3,634 & 6,095,521 \\
  Number of $\text{classes}^{\dagger}$   &     2 &       2 &       3 &     6 &     115 \\
  Columns of $X$ & 1 & 3 & 9 & 22 & 10,937 \\
      Size (compressed) & 3KiB & 21KiB & 145KiB & 2MiB & 36GiB \\
  Size (inflated) & 22KiB & 117KiB & 643KiB & 7MiB & 42GiB \\
  \bottomrule
  \end{tabular}
  \caption{Summary of tasks in \Mlz Task Suite (n=\ndatasets).
  \textsuperscript{$\dagger$}for classification tasks}
  \label{tab:ml-task-suite-summary}
\end{table}

\begin{table*}[t]
  \footnotesize
  \centering
  \setlength{\tabcolsep}{0.8\tabcolsep}
  \begin{tabularx}{\linewidth}{@{} lll >{\raggedright\arraybackslash}X @{}}
    \toprule
    Data Modality & Problem Type   & Tasks & Pipeline Template \\
    \midrule
    graph & community detection     & 2    & \mybox{CommunityBestPartition} \\
          & graph matching          & 9    & \mybox{link\_prediction\_feat\_extr} \mybox{graph\_feat\_extr} \mybox{CategoricalEncoder} \mybox{SimpleImputer} \mybox{StandardScaler} \mybox{XGBClassifier} \\
          & link prediction         & 1    & \mybox{link\_prediction\_feature\_extraction} \mybox{CategoricalEncoder} \mybox{SimpleImputer} \mybox{StandardScaler} \mybox{XGBClassifier} \\
          & vertex nomination       & 1    & \mybox{graph\_feature\_extraction} \mybox{categorical\_encoder} \mybox{SimpleImputer} \mybox{StandardScaler} \mybox{XGBClassifier} \\
    \greyrule
    image & classification          & 5    & \mybox{ClassEncoder} \mybox{preprocess\_input} \mybox{MobileNet} \mybox{XGBClassifier} \mybox{ClassDecoder} \\
          & regression              & 1    & \mybox{preprocess\_input} \mybox{MobileNet} \mybox{XGBRegressor} \\
    \greyrule
    multi table & classification    & 6    & \mybox{ClassEncoder} \mybox{dfs} \mybox{SimpleImputer} \mybox{StandardScaler} \mybox{XGBClassifier} \mybox{ClassDecoder} \\
          & regression              & 7    & \mybox{dfs} \mybox{SimpleImputer} \mybox{StandardScaler} \mybox{XGBRegressor} \\
    \greyrule
    single table & classification   & 234  & \mybox{ClassEncoder} \mybox{dfs} \mybox{SimpleImputer} \mybox{StandardScaler} \mybox{XGBClassifier} \mybox{ClassDecoder} \\
          & collaborative filtering & 4    & \mybox{dfs} \mybox{LightFM} \\
          & regression              & 87   & \mybox{dfs} \mybox{SimpleImputer} \mybox{StandardScaler} \mybox{XGBRegressor} \\
          & timeseries forecasting  & 35   & \mybox{dfs} \mybox{SimpleImputer} \mybox{StandardScaler} \mybox{XGBRegressor} \\
    \greyrule
    text  & classification          & 18   & \mybox{UniqueCounter} \mybox{TextCleaner} \mybox{VocabularyCounter} \mybox{Tokenizer} \mybox{pad\_sequences} \mybox{LSTMTextClassifier} \\
          & regression               & 9    & \mybox{StringVectorizer} \mybox{SimpleImputer} \mybox{XGBRegressor} \\
    \greyrule
    timeseries & classification     & 37   & \mybox{ClassEncoder} \mybox{dfs} \mybox{StandardImputer} \mybox{StandardScaler} \mybox{XGBClassifier} \mybox{ClassDecoder} \\
    \bottomrule
  \end{tabularx}
  \caption{\mltt{s} (data modality and problem type pairs) and associated \mlt{s} counts in the \Mlz Task Suite, along with default templates from \autobazaar (i.e., where we have curated appropriate \tem{s} to solve a task).}
  \label{tab:task-type-count-template}
\end{table*}

We made every effort to curate a corpus that was evenly balanced across \mltt{s}. Unfortunately, in practice, available datasets are heavily skewed to traditional ML problems of single-table classification and our task suite reflects this deficiency (though \pctnotsingletableclassification are not single-table classification). Indeed, among other evaluation suites, the OpenML 100 and the AutoML Benchmark \cite{bischl2017openml, Gijsbers2019} are both exclusively comprised of single-table classification problems. Similarly, evaluation approaches for \automl methods usually target the black-box optimization aspect in isolation \cite{Golovin2017, Guyon2015, Dewancker2016} without considering the larger context  of an end-to-end \ppl.

\subsection{Pipeline search}
\label{sec:evaluation:overall-performance}

We run the search process for all tasks in parallel on a heterogenous cluster of \nnodes AWS EC2 nodes. Each \mlt is solved independently on a node of its own over a \timelimit-hour time limit. Metadata and fine-grained details about every \ppl evaluated are stored in a MongoDB document store. The best \ppl{s} for each task, after checkpoints at 10, 30, 60, and 120 minutes of search, are selected by considering the cross-validation score on the training set and are then re-scored on the held-out test set.\footnote{Exact replication files and detailed instructions for the experiments in this section are included here: \url{\replicationurl} and can be further analyzed using our {\footnotesize\piex} library.}

\subsection{Computational bottlenecks}

\begin{figure}
      \includegraphics[width=\linewidth]{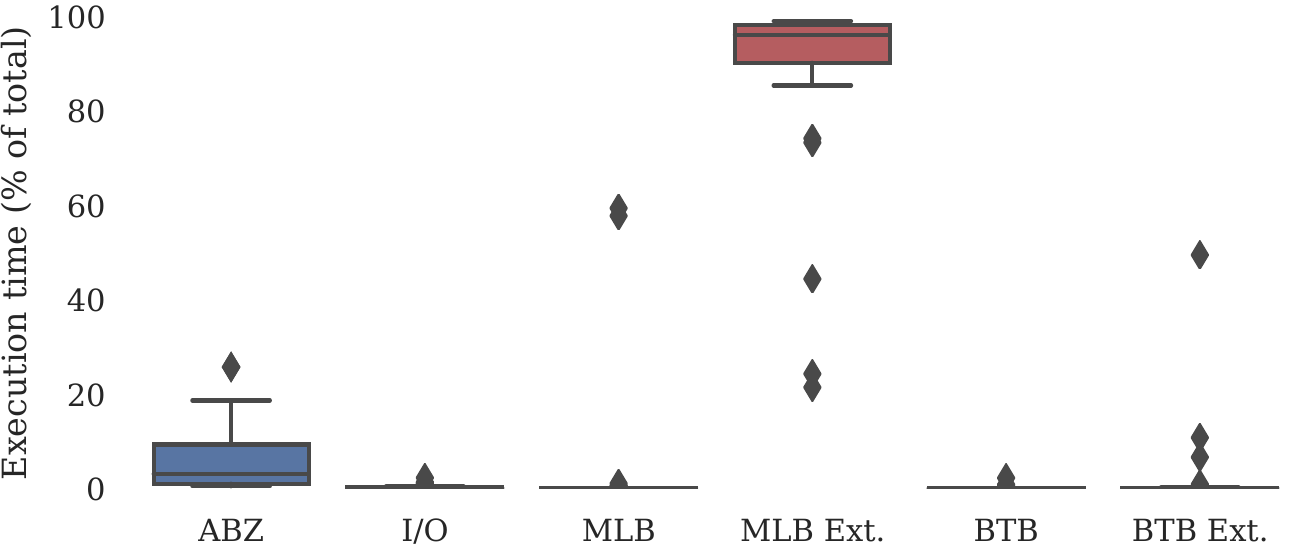}
      \caption{Execution time of \autobazaar \ppl search attributable to different libraries/components. The box plot shows quartiles of the distribution, 1.5$\times$ IQR, and outliers. MLB Ext and BTB Ext refer to calls to external libraries providing underlying implementations, like the scikit-learn \mytt{GaussianProcessRegressor} used in the \mytt{GP-EI} tuner. The vast majority of execution time is attributed to the underlying \pri{s} implemented in external libraries.}
      \label{fig:execution-time}
\end{figure}

We first evaluate the computational bottlenecks of the \autobazaar system. To assess this, we instrument \autobazaar and our framework libraries (\mlb, \mlp, \btb) to determine what portion of overall execution time for \ppl search is due to our runtime libraries vs. other factors such as I/O and underlying component implementation. The results are shown in \Cref{fig:execution-time}. Overall, the vast majority of execution time is due to execution of the underlying primitives (p25=90.2\%, p50=96.2\%, p75=98.3\%). A smaller portion is due to the \autobazaar runtime (p50=3.1\%) and a negligible (p50\textless0.1\%) portion of execution time is due to our other framework libraries and I/O. Thus, performance of \ppl execution/search is largely limited by the performance of the underlying physical implementation from the external library.

\subsection{\automl performance}

\begin{figure}[t]
      \includegraphics[width=0.8\linewidth]{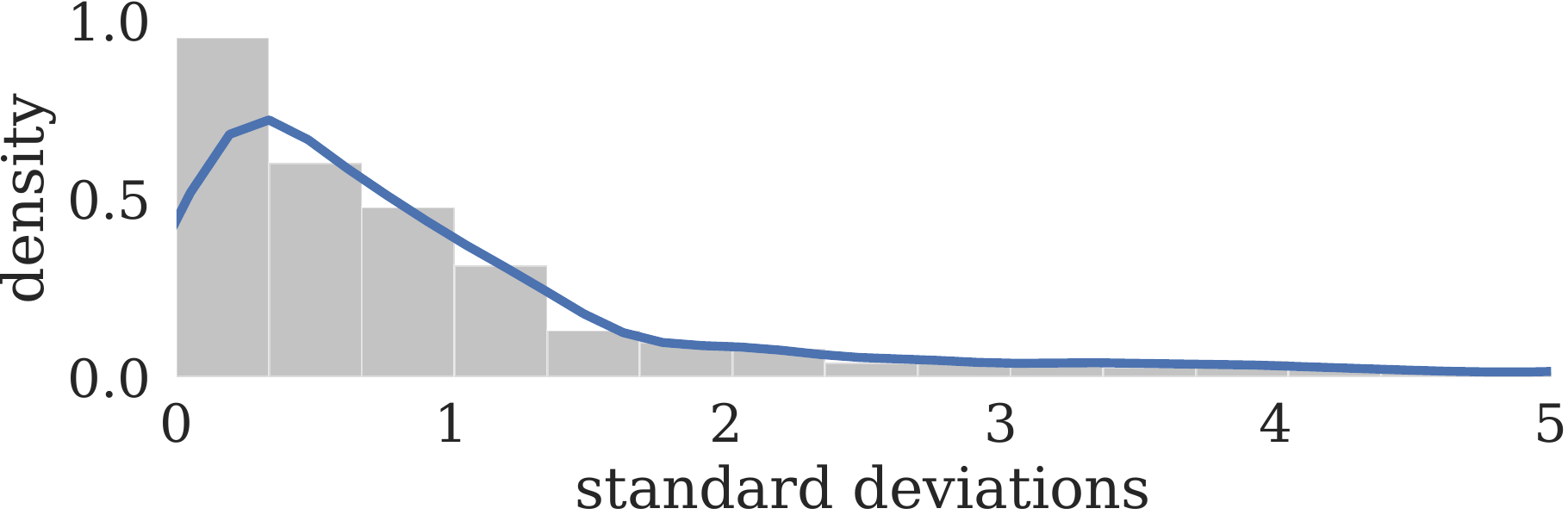}
      \caption{Distribution of task performance improvement from \Mlz \automl. Improvement for each task is measured as the score of the best \ppl less the score of the initial default \ppl, in standard deviations of all \ppl{s} evaluated for that task.}
      \label{fig:mlz_improvement_vs_default}
\end{figure}

One important attribute of \autobazaar is the ability to improve \ppl{s} for different tasks through tuning and selection. We measure the improvement in the best pipeline per task, finding that the average task improves its best score by \improvementinsd standard deviations over the course of tuning, and that \pctimprovedbyonesd of tasks improve by more than 1 standard deviation (\Cref{fig:mlz_improvement_vs_default}). This demonstrates the \autobazaar \ppl search effectiveness that a user may expect to obtain. However, as we describe in \Cref{sec:automl}, there are many possible \automl \pri{s} that can be implemented using our \tuner/\selector \api{s}; a comprehensive comparison is beyond the scope of our work.

\subsection{Expressiveness of \Mlz}
\label{sec:evaluation:expressiveness}

To further examine the expressiveness of \Mlz to solve a wide variety of tasks, we randomly selected 23 Kaggle competitions from 2018, comprising tasks ranging from image and time series classification to object detection and multi-table regression. For each task, we attempted to develop a solution using existing \pri{s} and atalogs.

Overall, we were able to immediately solve 11 tasks. We did not currently support 4 task types: image matching (2 tasks), object detection within images (4 tasks), multi-label classification (1 task), and video classification (1 task). We could readily support these within our framework by developing new \pri{s} and \ppl{s}. In the remaining tasks, multiple data modalities were provided to participants (\ie some combination of image, text, and tabular data). To support these tasks, we would need to develop a new ``glue'' primitive for concatenating separately-featurized data from each resources to create a single feature matrix. Though our evaluation suite contains many examples of tasks with multiple data resources of different modalities, we had written \ppl{s} customized to operate on certain common subsets (\ie tabular + graph).
We can never expect to have {\it already} implemented \ppl{s} for the innumerable diversity of ML task types, but we can still write new \pri{s} and \ppl{s} using our framework to solve these problems.

\subsection{Case study: evaluating ML primitives}
\label{sec:evaluation:case-study-evaluating-ml-primitives}

When new \pri{s} are contributed by the ML community, they can be incorporated into \tem{s} and \htem{s}, either to replace similar \pplstep{s} or to form the basis of new topologies. By running the end-to-end system on our evaluation suite, we can assess the impact of the \pri on general-purpose ML workloads (rather than overfit baselines).

In this first case study, we compare two similar \pri{s}: annotations for the XGBoost (\mytt{XGB}) and random forest (\mytt{RF}) classifiers. We ran two experiments, one in which \mytt{RF} is used in \tem{s} and one in which \mytt{XGB} is substituted instead.

We consider \totalrfxgbpipelines relevant \ppl{s} to determine the best scores produced for \ntasksxgbcomparison tasks. We find that the \mytt{XGB} \ppl{s} substantially outperformed the \mytt{RF} \ppl{s}, winning \pctxgbwins of the comparisons. This confirms the experience of practitioners, who widely report that XGBoost is one of the most powerful ML methods for classification and regression.

\subsection{Case study: evaluating \automl primitives}
\label{sec:evaluation:case-study-evaluating-automl-primitives}

The design of the \Mlz \automl system and our extensive evaluation corpus allows us to easily swap in new \automl \pri{s} (\Cref{sec:automl:automl-primitives}) to see to what extent changes in components like \tuner{s} and \selector{s} can improve performance in general settings.

In this case study, we revisit \cite{snoek2012practical}, which was influential for bringing about widespread use of Bayesian optimization for tuning ML models in practice. Their contributions include: (C1) proposing the usage of the \matern 5/2 kernel for \tuner meta-model (see \Cref{sec:automl:automl-primitives:tuners}), (C2) describing an integrated acquisition function that integrates over uncertainty in the GP \hyp{s}, (C3) incorporating a cost model into an EI per second acquisition function, and (C4) explicitly modeling pending parallel trials. How important was each of these contributions to a resulting tuner?

Using \Mlz, we show how a more thorough \textit{ablation study} \cite{lipton2018troubling}, not present in \cite{snoek2012practical}, would be conducted to address these questions, by assessing the performance of our general-purpose \automl system using different combinations of these four contributions. Here we focus on contribution C1.
We run experiments using a baseline \tuner with a squared exponential kernel (\mytt{GP-SE-EI}) and compare it with a \tuner using the \matern 5/2 kernel (\mytt{GP-Matern52-EI}). In both cases, kernel \hyp{s} are set by optimizing the marginal likelihood. Thus we can isolate the contributions of the proposed kernel in the context of general-purpose ML workloads.

In total, \totalmaternsepipelines \ppl{s} were evaluated to find the best \ppl{s} for a subset of \ntasksmaterncomparison tasks. We find that there is no improvement from using the \matern 5/2 kernel over the SE kernel --- in fact, the \mytt{GP-SE-EI} \tuner outperforms, winning \pctsqexpwins of the comparisons. One possible explanation for this negative result is that the \matern kernel is sensitive to \hyp{s} which are set more effectively by optimization of the integrated acquisition function. This is supported by the over-performance of the tuner using the integrated acquisition function in the original work; however, the integrated acquisition function is not tested with the baseline SE kernel, and more study is needed.

\section{Related work}
\label{sec:related}

Researchers have developed numerous algorithmic and software innovations to make it possible to create ML and \automl systems in the first place.

\paragraph{ML libraries}

High-quality ML libraries have originated over a period of decades. For general ML applications, scikit-learn implements many different algorithms using a common \api centered on the influential \mytt{fit}/\mytt{predict} paradigm \cite{sklearn_api}. For specialized analysis, libraries have been developed in separate academic communities, often with different and incompatible \api{s} \cite{opencv,networkX,lightfm,Kanter2015deep,nltk, tensorflow2015}. In \Mlz, we connect and link components of these libraries, only creating missing functionality ourselves.

\paragraph{ML systems}

Prior work has provided several approaches for making it easier to develop ML systems. For example, caret \cite{caret} standardizes interfaces and provides utilities for the R ecosystem, but without enabling more complex pipelines.
Recent systems have attempted to provide graphical interfaces, like \cite{Gong2019} and Azure Machine Learning Studio. Development of ML systems is closely tied to the execution environments needed to train, deploy, and update the resulting models. In SystemML \cite{boehm2016systemml} and Weld \cite{palkar2018evaluating}, implementations of specific ML algorithms are optimized for specific runtimes. Velox \cite{crankshaw2015missing} is an analytics stack component that efficiently serves predictions and manages model updates.

\paragraph{\automl libraries}

\automl research has often been limited to solving sub-problems of an end-to-end ML workflow, such as data cleaning \cite{Deng2017}, feature engineering \cite{Kanter2015deep, khurana2016cognito}, \hyp tuning \cite{snoek2012practical, gomes2012combining, Thornton2013AutoWEKACS, feurer2015efficient, Olson2016EvaluationOA, jamieson2016nonstochastic, li2017hyperband, hirzel2019lale}, or algorithm selection \cite{vanrijn2015fast, hirzel2019lale}. Thus \automl solutions are often not widely applicable or deployed in practice without human support. In contrast, \Mlz integrates many of these existing approaches and designs one coherent and configurable structure for joint tuning and selection of end-to-end \ppl{s}.

\paragraph{\automl systems}

These \automl libraries, if deployed, are typically one component within a larger system that aims to manage several practical aspects such as parallel and distributed training, tuning, and model storage, and even serving, deployment, and graphical interfaces for model building. These include ATM \cite{swearingen2017atm}, Vizier \cite{Golovin2017}, and Rafiki \cite{Wang2018}, as well as commercial platforms like Google AutoML, DataRobot, and Azure Machine Learning Studio. While these systems provide many benefits, they have several limitations. First, they each focus on a specific subset of ML use cases, such as computer vision, NLP, forecasting, or hyperparameter tuning. Second, these systems are designed as proprietary applications and do not support community-driven integration of new innovations. \Mlz provides a new approach to \textit{developing} such systems in the first place: it supports a wide variety of \mltt{s}, and builds on top of a community-driven ecosystem of ML innovations. Indeed, it could serve as the backend for such ML services or platforms.

The DARPA D3M program \cite{lippmann2016d3m}, of which we are participants, aims to spur development of automated systems for model discovery for use by non-experts. Several differing approaches are being developed within this context. For example, Alpine Meadow \cite{shang2019democratizing} focuses on efficient search for producing interpretable ML pipelines with low latencies for interactive usage. It combines existing techniques from query optimization, Bayesian optimization, and multi-armed bandits to efficiently search for pipelines. AlphaD3M \cite{drori2018alphad3m} formulates a pipeline synthesis problem and uses reinforcement learning to construct pipelines. In contrast, \Mlz is a framework to develop ML or \automl systems in the first place. While we present our open-source \autobazaar system, it is not the primary focus of our work and represents a single point in the design space of \automl systems using our framework libraries. Indeed, one could use specific \automl approaches like the ones described by Alpine Meadow or AlphaD3M for pipeline search within our own framework.

\section{Conclusion}
\label{sec:conclusion}

Throughout this paper, we have built up abstractions, interfaces, and software components for data scientists, data engineers, and other practitioners to effectively develop machine learning systems. Developers can use \Mlz to compose one-off \ppl{s}, tunable \tem{s}, or full-fledged \automl systems. Researchers can contribute individual ML or \automl \pri{s} and make them easily accessible to a broad base for inclusion in end-to-end solutions.

We have applied this approach to several real-world ML problems and entered our \automl system in an automated modeling program. As we collect more and more scored \ppl{s}, we expect opportunities will emerge for meta-learning and debugging on \mlt{s} and \ppl{s}, as well as the ability to track progress and transfer knowledge within data science organizations. We will focus on several complementary extensions in future work. These include continuing to improve our \automl system and making it more robust for everyday use by a diverse user base, and studying how to best support users of different backgrounds in using and interacting with ML and \automl systems.

\begin{acks}
  The authors would like to thank Plamen Valentinov Kolev for contributions running experiments and testing datasets. The authors also acknowledge the contributions of the following people: Laura Gustafson, William Xue, Akshay Ravikumar, Ihssan Tinawi, Alexander Geiger, Sarah Alnegheimish, Saman Amarasinghe, Stefanie Jegelka, Zi Wang, Benjamin Schreck, Seth Rothschild, Manual Alvarez Campo, Sebastian Mir Peral, Peter Fontana, and Brian Sandberg. The authors are part of the DARPA Data-Driven Discovery of Models (D3M) program, and would like to thank the D3M community for the discussions around the design. This material is based on research sponsored by DARPA and Air Force Research Laboratory (AFRL) under agreement number FA8750-17-2-0126. The U.S. Government is authorized to reproduce and distribute reprints for Government purposes notwithstanding any copyright notation thereon. The view and conclusions contained herein are those of the authors and should not be interpreted as necessarily representing the official policies of endorsements, either expressed or implied, of DARPA and Air Force Research Laboratory (AFRL) or the U.S. Government. Authors MJS, CS, and KV acknowledge support and user feedback from Iberdrola S.A. and SES S.A.

\end{acks}

\bibliographystyle{ACM-Reference-Format}

\bibliography{references.bib}

\end{document}